\documentclass[aps,prd,twocolumn,nofootinbib,superscriptaddress]{revtex4}
\usepackage[hypertex]{hyperref}
\usepackage{graphicx}
\def\lesssim{\mathrel{\mathpalette\vereq<}}
\def\gtrsim{\mathrel{\mathpalette\vereq>}}

\newcommand{\be}{\begin{equation}}
\newcommand{\ee}{\end{equation}}
\newcommand{\bea}{\begin{eqnarray}}
\newcommand{\eea}{\end{eqnarray}}

\begin{document}

\pagestyle{plain}

\title{\begin{flushright} \texttt{UMD-PP-08-028} \end{flushright}
Neutrino Mass Seesaw at the Weak Scale, the Baryon Asymmetry, and the LHC}

\author{Steve Blanchet}
\author{Z. Chacko}
\author{Rabindra N. Mohapatra}
\affiliation{Department of Physics, University of Maryland, College Park, MD, 20742}


\begin{abstract}

We consider theories where the Standard Model (SM) neutrinos acquire masses through the seesaw mechanism at the weak
scale. We show that in such a scenario, the requirement that any pre-existing baryon asymmetry, regardless of its origin,
not be washed out leads to correlations between the pattern of SM neutrino masses and the spectrum of new particles at
the weak scale, leading to definite predictions for the LHC. For type I seesaw models with a TeV scale $Z'$ coupled to SM
neutrinos, we find that for a normal neutrino mass hierarchy, at least one of the right-handed neutrinos must be
`electrophobic', decaying with a strong preference into final states with muons and tauons rather than electrons. For
inverted or quasi-degenerate mass patterns, on the other hand, we find upper bounds on the mass of at least one
right-handed neutrino. In particular, for an inverted mass hierarchy, this bound is 1 TeV, while the corresponding upper
limit in the quasi-degenerate case is 300 GeV. Similar results hold in type III seesaw models, albeit with somewhat more
stringent bounds. For the Type II seesaw case with a weak scale $SU(2)$ triplet Higgs, we again find that an interesting
range of Higgs triplet masses is disallowed by these considerations.

\end{abstract}

 \maketitle


\section{Introduction}

Neutrino masses constitute concrete evidence for the existence of physics beyond the Standard Model (SM). However, much
about the neutrino sector remains to be understood. It is not known whether neutrinos are Dirac or Majorana, or whether
the pattern of neutrino masses is hierarchical, inverse hierarchical or quasi-degenerate. Further, the dynamics by which
the neutrinos acquire their masses, and the scale at which this occurs, are also not known.

 One interesting possibility is that the SM neutrinos acquire their masses at or close to the weak
 scale through the seesaw mechanism~\cite{TypeI}. In such a scenario the
dynamics underlying
 neutrino mass generation might be accessible to the Large Hadron
Collider (LHC)~\cite{smir}. This could happen
 in several different ways. For example, if the SM neutrinos are charged
under a new $U(1)$ gauge
 symmetry that is broken only close to the weak scale, they are prevented
from acquiring mass at
 higher scales. In these theories a Type I seesaw at the weak scale is
perhaps the simplest
 possibility for neutrino mass generation, with some of the
additional particles required for
 cancellation of the $[U(1)]^3$ anomalies playing the role of
right-handed (RH) neutrinos. Such a
 scenario is promising for the LHC because the RH neutrinos
could be pair-produced
 through decays of the $Z'$ gauge boson, and are in general
straightforward to detect through their
 lepton-number-violating decays. Other classes of seesaw models that are
exciting for the LHC are
 Type II models with an $SU(2)$ triplet Higgs at the weak
scale~\cite{TypeII}, and Type
III models with $SU(2)$
triplet RH neutrinos at the weak scale~\cite{TypeIII}.

In this letter we show that in theories where the Standard Model (SM) neutrinos acquire Majorana masses through the
seesaw mechanism at the weak scale, the requirement that any pre-existing baryon asymmetry is not almost entirely washed
out can be used to correlate the pattern of SM neutrino masses to the spectrum of new particles at the weak scale. Then,
unless the baryon asymmetry is generated at or below the weak scale, or the pre-existing baryon asymmetry is extremely
large, this leads to definite predictions at the LHC for each of the three classes of seesaw models described above.
\begin{itemize}
\item
In the case of Type I seesaw models with a weak-scale $Z'$, if the SM neutrinos are hierarchical, at least one of the
RH neutrinos must be `electrophobic', decaying with a strong preference into final states with muons and tauons
rather than electrons. On the other hand, if the SM neutrinos exhibit an inverted (quasi-degenerate) pattern of masses,
at least one of the RH neutrinos must be lighter than 1 TeV (300 GeV).
\item
For Type III seesaw models, the corresponding upper bound for $SU(2)$ triplet RH neutrinos is 300~GeV for inverted neutrino
mass pattern and 170~GeV for the quasi-degenerate case.
\item
In the case of Type II seesaw models, a significant portion of the triplet VEV and mass
($v_\Delta$,$M_\Delta$) parameter space accessible at the LHC is disfavored.
\end{itemize}
It should be noted that our objective is different from that of previous works
whose purpose was to explain the matter-antimatter
asymmetry thanks to leptogenesis (for a recent review, see~\cite{Review}).
Leptogenesis in the Type I case has been
extensively studied, and it was found that there exists a lower bound on
the scale of leptogenesis around $10^9$~GeV in the
limit of hierarchical RH neutrinos~\cite{DavidsonIbarra,Review1,Giudice}.
In the Type II case, \cite{HambyeII} found that
there is also a lower bound which points to a very high scale. Similarly,
the Type III was worked out in \cite{HambyeIII},
where a similar bound was obtained. Let us stress here that these bounds
are in no way inconsistent with our approach,
which is not concerned in any way with the mechanism whereby the baryon
asymmetry was originally generated. Instead we
study the constraints that arise from requiring that any such asymmetry
survive to low temperatures, and is not erased at
the weak scale. More precisely, in this work we focus exclusively on
washout, assuming that the effects responsible for the
generation of the asymmetry are negligible at the weak scale.

To see the origin of the connection between the properties of the sub-TeV scale fields responsible for neutrino mass
generation (i.e. singlet or triplet RH neutrinos or Higgs triplet fields) and the baryon asymmetry, note that since
Majorana neutrino masses necessarily violate lepton number, any mechanism whereby the SM neutrinos acquire Majorana
masses at the weak scale will generally lead to some amount of lepton number violation at low energies. If lepton-number-violating processes are in equilibrium at the weak scale, they can, in combination with sphalerons~\cite{sphalerons},
erase any pre-existing baryon asymmetry, regardless of how the baryon asymmetry originated.

As a concrete example, consider a toy Type I model with a single SM lepton doublet $L$ that acquires a Majorana mass
after coupling through the Higgs $H$ to a single RH neutrino $N$ that has a weak-scale mass $M$. The relevant part of the
Lagrangian is
\begin{equation}
\lambda L H N + M N^2 \; + \; {\rm h.c.}
\end{equation}
At temperatures $T \geq M$ the rate for lepton-number-violating decays and inverse decays of $N$
is given by
\begin{equation}
\Gamma \sim \lambda^2 \frac{M^2}{T} \sim m_{\nu} \frac{M^3}{v^2 T}
\end{equation}
where $v = \langle H \rangle$ is the SM Higgs VEV and $m_{\nu} \sim \lambda^2 v^2/M$ is the SM
neutrino mass. For this process to be in equilibrium, this rate must be larger than the expansion
rate $H \sim T^2/M_{\rm Pl}$, where $M_{\rm Pl}$ is the Planck scale. Lepton-number-violating
decays will be in equilibrium for $T \sim M$ if $m_{\nu} \gtrsim v^2/M_{\rm Pl}$. For $m_{\nu}$ of
order 0.05~eV, the atmospheric scale, this condition is satisfied in our toy model, and any
pre-existing baryon asymmetry will endure washout.
However, whether or not the baryon asymmetry is erased in realistic seesaw models is a far more
detailed question. There are several reasons for this.
\begin{itemize}
\item
Since the medium distinguishes all three lepton flavors $L_e$,
$L_{\mu}$ and  $L_{\tau}$ at the weak scale, and since sphalerons conserve
$B/3-L_{\alpha}$, ($\alpha~=~e,\mu,\tau$) lepton flavor violating processes must be in equilibrium for each lepton
flavor
in order for the baryon asymmetry to be efficiently erased. This requires taking
into account the full spectrum of neutrino masses and mixings.
\item
The extent to which the baryon asymmetry is erased depends on the period of time that the
lepton-flavor-violating processes, most importantly inverse decays, are in equilibrium.
For $m_{\nu}$ of order the atmospheric
scale, these processes are only in equilibrium for a short time. It then requires a careful
analysis to determine how much of the original baryon asymmetry survives. The relevant Boltzmann equation
describing the washout process by the seesaw mediator(s) of mass $M$ can be written in the general form
\begin{equation}\label{Boltz}
{{\rm d} Y_{B/3-L_{\alpha}}\over {\rm d}z}=- W_{\alpha}(z) Y_{B/3-L_{\alpha}},
\end{equation}
where $z=M/T$, $Y_{B/3-L_{\alpha}}$ is the $B/3-L_{\alpha}$ number density over the
entropy density, and $W_{\alpha}$ is a generic washout term which will have to be specified
in each particular case. This formulation of the washout term is consistent with leptogenesis analyses
in the Type I and III cases~\cite{Review,HambyeIII} by setting the $C\!P$ asymmetry to zero. In the
Type II case~\cite{HambyeII}, the situation is slightly more complicated since the scalar triplets carry
non-zero hypercharge. As we will show explicitly in Section~\ref{sec:typeII}, in that case,
the evolution of the scalar triplet asymmetry must be tracked as well. However, we will find
that the more complicated system of equations reduces to this simple form in a certain limit.

The solution to Eq.~(\ref{Boltz}) can be easily obtained
\begin{equation}\label{analsol}
Y_{B/3-L_{\alpha}}(z)=Y_{B/3-L_{\alpha}}^{\rm in} \exp\left[-\int_{z_{\rm in}}^z {\rm d}z'\,
W_{\alpha}(z')\right],
\end{equation}
where $Y_{B/3-L_{\alpha}}^{\rm in}$ stands for any pre-existing asymmetry at $z_{\rm in}\ll 0.1$.
One can therefore see that any pre-existing asymmetry will be \emph{exponentially} washed out
provided the integral is large.
\item
The baryon asymmetry, having been erased, can in principle be regenerated by late out-of-equilibrium decays of RH
neutrinos. However, if the masses of the RH neutrinos are of order the weak scale as in the case of interest, the
relevant Yukawa couplings are generally much too small to allow a large-enough asymmetry to be generated.
However, an exception to this rule occurs if the the RH neutrinos are extremely degenerate, at the level of one part in
$10^{10}$, in which case resonant leptogenesis can occur~\cite{Pilaftsis1}. In what follows we will assume that the RH neutrinos are not
sufficiently degenerate to allow this possibility, leaving this very special case for future work.
\end{itemize}

In the subsequent sections we analyse each of the three classes of seesaw models in turn, and determine the precise
conditions under which the baryon asymmetry is erased, and the consequences for LHC phenomenology. We shall place limits
on the spectrum of particles at the weak scale by requiring that the baryon asymmetry not be washed out by a factor
greater than $10^{6}$. It is important to note that these bounds can be evaded if the baryon asymmetry is generated
below the weak scale, or alternatively, if the primordial baryon asymmetry generated at high scales is extremely large,
of order $10^{-3}$ or more. However, such a large primordial asymmetry cannot be generated by the familiar
out-of-equilibrium decays of a heavy field, but instead requires a very efficient mechanism of baryogenesis, such as
the one proposed by Affleck and Dine~\cite{affleck}. If the LHC were to discover that these bounds are
violated, it would shed new light on the
mechanism that generates the baryon asymmetry of the Universe.

\section{The Type I Seesaw }
For the Type I seesaw case, we assume the presence of three RH neutrinos
$N_i$, $i=1,2,3$ with masses $M_i$ around the weak scale.
The washout parameter introduced in Eq.~(\ref{Boltz})
is given in this case by
\begin{equation}
W^{\rm I}_{\alpha}(z)=\sum_i W^{\rm ID}_{i\alpha}(z_i),
\end{equation}
where $z_i=z M_i/M_1$. The washout is dominated by inverse decays $W^{\rm ID}$,
which can be conveniently expressed as~\cite{Review1}
\begin{equation}
W^{\rm ID}_{i\alpha}={1\over 4} K_{i\alpha} \mathcal{K}_1(z_i) z_i^3,
\end{equation}
where $\mathcal{K}_1(z)$ is the modified Bessel function, and
\begin{equation}\label{decpar}
K_{i\alpha}={\widetilde{\Gamma}_{\rm D} (N_i\to \ell_{\alpha} H +\bar{\ell}_{\alpha} H^{\dagger})\over
H(z_i=1)}={|\lambda_{\alpha i}|^2 v^2\over M_i m_{\star}},
\end{equation}
with $m_{\star}=1.08\times 10^{-3}$~eV.

\begin{figure}
\includegraphics[width=0.3\textwidth,angle=-90]{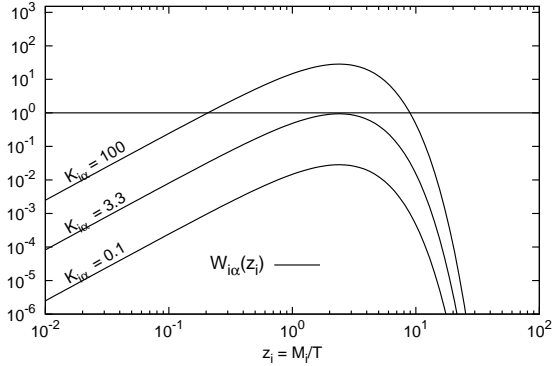}
\caption{Washout function $W_{i\alpha}(z_i)$ for different values of $K_{i\alpha}$. The area
below the curve determines the total washout [cf.~Eq.~(\ref{analsol})].}\label{fig:1}
\end{figure}

We show in Fig.~\ref{fig:1} how the function
$W_{i\alpha}(z_i)$ behaves for different values of $K_{i\alpha}$. Note that an identical figure
would be obtained in the Type II and III cases, where only the definition of $K_{i\alpha}$ is different.
It is useful to derive the maximum washout possible, which from Fig.~\ref{fig:1} is obtained for
$M\gtrsim 10 T$. In this limit ($z\gg 1$), the integral in Eq.~(\ref{analsol})
can be solved analytically~\cite{blanchet}:
\begin{equation}\label{washoutanal}
Y_{B/3-L_{\alpha}}(\infty)=Y_{B/3-L_{\alpha}}^{\rm in} \exp\left[-{3\pi\over 8}\sum_i K_{i\alpha}\right].
\end{equation}
However, more precisely, the upper bound on the integral in Eq.~(\ref{analsol}) should be set by the decoupling temperature
of the sphalerons. For $M_H=120$~GeV, it is given by $T_{\rm dec}\simeq 130$~GeV~\cite{Burnier}.
At fixed temperature $T=T_{\rm dec}$, it can be seen from Fig.~\ref{fig:1} that the area
below the curve $W(z)$ increases as $M$ increases, so that the washout increases
as well [see Eq.~(\ref{analsol})]. Therefore, by imposing that the washout does not exceed a
factor of $10^6$, we will obtain upper bounds on $M$.

The question then is: how to know the magnitude of $K_{i\alpha}$. This can be answered using the
parametrization of the Yukawa coupling matrix~\cite{CasasIbarra}
\begin{equation}
\lambda_{\alpha i}=\left(U \sqrt{D_m} \Omega \sqrt{D_M}\right)_{\alpha i}/v,
\end{equation}
where $U$ is the PMNS mixing matrix, $D_m$ and $D_M$ are diagonal matrices
for the masses of light and heavy neutrinos, respectively, and $\Omega$ is
a complex orthogonal matrix. Using this parametrization, one can
express the decay parameters in Eq.~(\ref{decpar}) in the following way:
\begin{equation}
K_{i\alpha}={1\over m_{\star}}\left|\sum_j \sqrt{m_j} U_{\alpha j}\Omega_{ji}\right|^2.
\end{equation}
It should be noted that $K_i\equiv \sum_{\alpha} K_{i\alpha}$ obey the following inequality:
\begin{equation}
\sum_i K_i \geq {\sum_i m_i\over m_{\star}}.
\end{equation}
Without flavor effects, the washout implied would be huge. With flavor effects the
washout is reduced, but one has to check to what extent.

In the case of normal hierarchy and $m_1 \ll m_{\rm sol}$, due to the small $U_{e3}$ entry in the PMNS matrix,
the washout in flavor $e$ is
typically suppressed, and it can be as low as $\sum_i K_{ie}\sim 2$, implying little washout.
Clearly, the washout in the other two flavors is very effective. We show below that in this case
at least one RH neutrino will decay electrophobically. Let us explain how this
comes about by first writing the branching ratio for the decay of $N_i$ into each
flavor $\alpha$:
\begin{equation}
B_{i\alpha}\equiv {\left|\sum_j \sqrt{m_j} U_{\alpha j}\Omega_{ji}\right|^2 \over
\sum_j m_j |\Omega_{ji}|^2}.
\end{equation}
In the case of normal hierarchy with $m_1\ll m_{\rm sol}$, for flavor $e$
the largest mass eigenstate $m_3=m_{\rm atm}$ is coupled to the
small entry $U_{e3}<0.2~(3\sigma)$. One can then envisage three
situations: $|\Omega_{3i}|\ll |\Omega_{2i}|$, $|\Omega_{3i}|\sim
|\Omega_{2i}|$, or $|\Omega_{3i}|\gg |\Omega_{2i}|$. In the first case,
the decay will be roughly 1/3 in each flavor, but the total rate
will be suppressed since it is driven by the solar scale. In the second
case, both terms are comparable, implying that $N_i$ will decay
typically only 8\% into $e$. In the third case, one obtains
$B_{ie}\simeq|U_{e3}|^2<4\%$. There is a however a loophole in the first
case. From the orthogonality of $\Omega$, $\sum_i \Omega_{ij}^2=1$,
it is not possible to have $|\Omega_{3i}|\ll |\Omega_{2i}|$ for
all $i$. Hence, for at least one RH neutrino, the
first case is excluded and the decay will be electrophobic.

\begin{figure}
\includegraphics[width=0.3\textwidth,angle=-90]{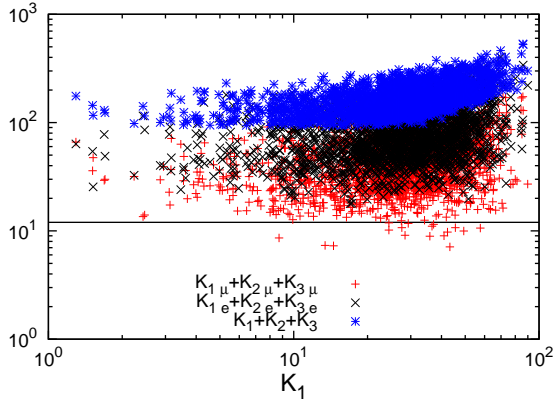}
\caption{Scatter plot of $\sum_i K_{ie}$ (green x's), $\sum_i K_{i\mu}$ (red crosses)
as well as $\sum_i K_i$ (blue stars) versus $K_1$
in the case of inverted hierarchy with $m_3=0$ varying the unknown parameters in the PMNS and
$\Omega$ matrices keeping $|\Omega|<1$.
We obtain that 99\% of the points lie above $\sum_i K_{i\alpha}=12$.
}\label{fig:scatter}
\end{figure}
For an inverted hierarchy and $m_3\ll m_{\rm sol}$, the situation is different because both $m_1$ and $m_2$ are roughly at the
atmospheric scale, implying typically a larger washout than in the previous case. Quantitatively, varying $0\leq
|\Omega_{ij}|\leq 1$ as well as the unknown phases in the PMNS phases, we find that 99\% of the points satisfy $\sum_i
K_{i\alpha}\geq 12$. In Fig.~\ref{fig:scatter} we show the scatter plot from which this lower bound was extracted.
This means that the baryon asymmetry
will be washed out by more than a factor of $10^{6}$ when $M_1>1$~TeV for
99\% of the parameter space. Thus we consider conservatively 1~TeV to be the upper bound on
the mass of the lightest RH neutrino.
Note that this result also holds when $|\Omega_{ij}|$ is allowed to take larger values, because the washout
increases as $K\propto |\Omega|^2$.

\begin{figure}
\includegraphics[width=0.5\textwidth]{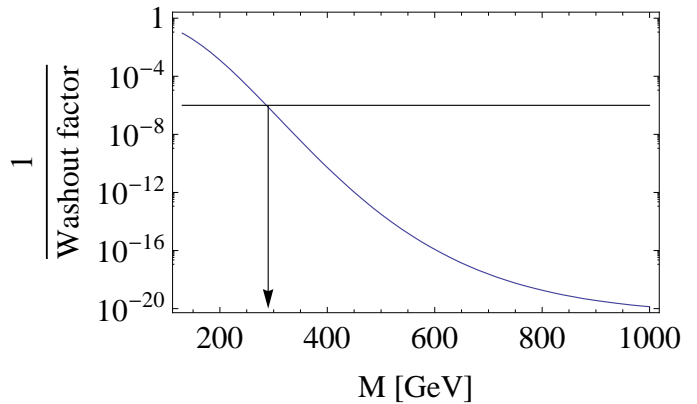}
\caption{Washout factor for the quasi-degenerate spectrum, fixing $\sum_i K_{i\alpha}= 40$. We find the upper bound
for a washout factor lower than $10^6$ to be around 300 GeV.
}\label{fig:washout}
\end{figure}

For a quasi-degenerate spectrum, $m_1\simeq m_2\simeq m_3\simeq 0.1$~eV, the washout is expected to be even larger. It
turns out that $\sum_i K_{i\alpha}\geq 40$, implying an upper bound $M_1<300$~GeV, a constraint
stronger than the inverted hierarchy case. We present in Fig.~\ref{fig:washout} a plot of the
washout factor in Eq.~(\ref{analsol}) versus the RH neutrino mass. This figure shows explicitly
how the upper bound of about 300~GeV was obtained.

\subsection{Neutrino textures}

We now comment on the possibility of evading the bounds derived
above in the case of inverted hierarchy or quasi-degenerate spectrum
by a suitable texture in the Yukawa coupling matrix for neutrinos.
The question is whether the points in the parameter space which we
did not consider (e.g. points for which $\sum_i K_{i\alpha}\ll 12$
in Fig.~\ref{fig:scatter}) require fine-tuning of model parameters
to get desired neutrino masses and mixings or not.
 Our conclusion
will be that the structure needed in order to avoid the washout is
not compatible with an inverted hierarchy or a quasi-degenerate
spectrum, except with a very careful tuning of the parameters.

The most minimal condition for evading our bounds in the
case of inverted or quasi-degenerate spectra is to reduce the
washout in a particular flavor, as given below:
\begin{eqnarray}
K&=&\left(\begin{array}{ccc}
|\epsilon_1|^2& |\epsilon_2|^2& |\epsilon_3|^2\\
|a_1|^2&|a_2|^2 & |a_3|^2\\
|b_1|^2&|b_2|^2 &|b_3|^2
\end{array}\right),
\left(\begin{array}{ccc}
|a_1|^2&|a_2|^2 & |a_3|^2\\
|\epsilon_1|^2& |\epsilon_2|^2& |\epsilon_3|^2\\
|b_1|^2&|b_2|^2 &|b_3|^2
\end{array}\right), \nonumber\\
&&\left(\begin{array}{ccc}
|a_1|^2&|a_2|^2 & |a_3|^2\\
|b_1|^2&|b_2|^2 &|b_3|^2\\
|\epsilon_1|^2& |\epsilon_2|^2& |\epsilon_3|^2
\end{array}\right)
\end{eqnarray}
where we impose that $\sum_i |\epsilon_i|^2 \ll \sum_i|a_i|^2, \sum_i|b_i|^2$. Let us now
calculate the form of the light neutrino mass matrices from the above $K$ matrix structure using Type I seesaw and
diagonal RH neutrino mass matrix. We obtain the following three different forms:
\begin{eqnarray}\label{mnuepsilon}
m_{\nu}&=&v^2\left(\lambda {1\over M} \lambda^T\right)\nonumber\\
&=&m_{\star} \left(\begin{array}{ccc} \mathcal{O}(\epsilon^2)
& \mathcal{O}(\epsilon) & \mathcal{O}(\epsilon)\\
\mathcal{O}(\epsilon)&A& C\\
\mathcal{O}(\epsilon)&C& B\end{array}\right),\\
 && m_{\star} \left(\begin{array}{ccc}
A&\mathcal{O}(\epsilon)& C\\
\mathcal{O}(\epsilon)& \mathcal{O}(\epsilon^2) & \mathcal{O}(\epsilon)\\
C&\mathcal{O}(\epsilon)& B\end{array}\right),\nonumber\\
&& m_{\star} \left(\begin{array}{ccc}
A& C&\mathcal{O}(\epsilon)\\
C& B&\mathcal{O}(\epsilon)\\
\mathcal{O}(\epsilon)
& \mathcal{O}(\epsilon) & \mathcal{O}(\epsilon^2)\end{array}\right),\nonumber
\end{eqnarray}
where
\begin{eqnarray}
A&=&a_1^2+a_2^2+a_3^2\\
B&=&{b_1^2}+{b_2^2}+{b_3^2}\\
C&=&{a_1 b_1}+{a_2 b_2}+{a_3 b_3}.
\end{eqnarray}
Due to the condition $\sum_i |\epsilon_i|^2 \ll \sum_i|a_i|^2, \sum_i|b_i|^2$, it can be
immediately seen that one eigenvalue will necessarily be suppressed,
and therefore it not consistent with a quasi-degenerate spectrum.

For the case of inverted hierarchy, it can be shown that the
condition to have one zero eigenvalue and two degenerate
eigenvalues, i.e. $m_3= 0$ and $m_1= m_2$, is given by $|A|=|B|$ and
$|C|=0$. Allowing for a small perturbation, we have a matrix of the
following form [for illustration we show only the first matrix in
Eq.~(\ref{mnuepsilon})]:
\begin{equation}
m_{\nu}= m_{\star} \left(\begin{array}{ccc} \mathcal{O}(\epsilon^2)
& \mathcal{O}(\epsilon) & \mathcal{O}(\epsilon)\\
\mathcal{O}(\epsilon)&A & \mathcal{O}(\epsilon)\\
\mathcal{O}(\epsilon)&\mathcal{O}(\epsilon) &
A+\mathcal{O}(\epsilon)
\end{array}\right)
\end{equation}
Note that the potential phase difference between $A$ and $B$ can be
absorbed by making the transformation $\nu_3\to {\rm e}^{{\rm
i}\phi}\nu_3$. Diagonalizing a mass matrix of this form, it is easy
to show that one mixing angle, $\theta_{23}$, can be large, whereas
the other two, $\theta_{12}$ and $\theta_{13}$ have to be small
because of the smallness of $\epsilon$. This is at odds
with the observed large solar mixing angle, which, though not
maximal, is large.

\subsubsection{Examples of Neutrino Mass Models}
It seems from the previous discussion that the observed neutrino
masses and mixings constrain the possible patterns of the $K$ matrix
in such a way that a small washout in one flavor is not expected for
inverted and quasi-degenerate spectra. In order to further illustrate this
point, we give two examples of known neutrino models, which lead to
different neutrino mass hierarchies.

\paragraph{$L_e-L_{\mu}-L_{\tau}$ symmetry case:} The well-known symmetry
$L_e-L_{\mu}-L_{\tau}$~\cite{Petcov:1982ya} predicts an inverted
spectrum. One way to implement this symmetry is to use $3\times 2$
seesaw with the following Yukawa coupling matrix:
\begin{equation}
\lambda=\left(\begin{array}{cc}
a_1 & \mathcal{O}(\epsilon)  \\
 \mathcal{O}(\epsilon)& a_2 \\
 \mathcal{O}(\epsilon) &a_3  \end{array}\right).
\end{equation}
The $\mathcal{O}(\epsilon)$ terms denote small breaking of this symmetry.
The $L_e-L_{\mu}-L_{\tau}$ charges of $N_{e,\mu}$ are $(+1, -1)$
respectively. The RH neutrino mass matrix in the flavor basis is
given by
\begin{equation}
\left(\begin{array}{cc}
0 & M  \\ M& 0  \end{array}\right).
\end{equation}
In the basis where the RH neutrino masses are diagonal, it can be immediately
seen that the $K$ matrix,
\begin{equation}
K={v^2\over \sqrt{2}M m_{\star}}\left(\begin{array}{cc}
|a_1|^2 & |a_1|^2  \\
 |a_2|^2& |a_2|^2 \\
 |a_3|^2 & |a_3|^2  \end{array}\right),
\end{equation}
does not allow the washout to be avoided
in any flavor.

One can also have examples with $3\times 3$ seesaw with obvious
assignments under $L_e-L_{\mu}-L_{\tau}$ symmetry for the lepton
doublet and singlet fields. In the strict symmetry limit, the RH
neutrino mass matrix is singular. One can therefore consider a
softly broken $L_e-L_{\mu}-L_{\tau}$ symmetry with a RH neutrino
mass matrix of the following form:
\begin{equation}\label{MajoLe}
\left(\begin{array}{ccc} 0 & M_1 & M_2  \\ M_1& 0 & 0 \\ M_2 & 0
& M_3
\end{array}\right).
\end{equation}
where the $M_3$ term breaks the symmetry softly. The Yukawa matrix 
in this case is:
\begin{equation}\label{YukLe}
\lambda=\left(\begin{array}{ccc} a & 0 &0 \\ 0& b & c \\ 0 & d & e
\end{array}\right).
\end{equation}
Using seesaw in the limit $M_3\gg M_{1,2}$, one finds the following form for the
neutrino mass matrix, which is characteristic of the inverted hierarchy:
\begin{equation}
m_{\nu}=v^2\left(\begin{array}{ccc} 0 & ab/M_1 & ad/M_1 \\ ab/M_1 &
\mathcal{O}(\epsilon)&\mathcal{O}(\epsilon)\\ ad/M_1 & \mathcal{O}(\epsilon) & \mathcal{O}(\epsilon)
\end{array}\right).
\end{equation}
where $\epsilon \sim 1/{M_3}$. From Eqs.~(\ref{MajoLe}) and (\ref{YukLe}) we obtain
the following $K$ matrix:
\begin{equation}
K={v^2\over \sqrt{2} m_{\star}}\left(\begin{array}{ccc}
|a|^2/M_1 & |a|^2/M_2 & \mathcal{O}(\epsilon)  \\
 |b|^2/M_1& |b|^2/M_2 & \mathcal{O}(\epsilon) \\
 |d|^2/M_1 & |d|^2/M_2& \mathcal{O}(\epsilon)  \end{array}\right),
\end{equation}
which clearly does not allow for a weak washout in any flavor.

\paragraph{Quasi-degenerate $A_4$ model:} The above conclusion holds also in models that
predict quasi-degenerate neutrinos. As an example, consider a
model based on discrete non-Abelian $A_4$ symmetry~\cite{Babu:2002dz}. 
The Yukawa matrix in this model is given by
\begin{equation}
\lambda={1\over \sqrt{3}}\left(\begin{array}{ccc}
1 & 1& 1 \\
1 & \omega  & \omega^2\\
1 & \omega^2  & \omega\end{array}\right),
\end{equation}
where $\omega$ is a phase. Since the moduli
of all Yukawa couplings are of the same order, any spectrum of RH neutrino masses 
will lead to a strong washout in all flavors.

In summary, realistic neutrino mass models do not lead to dramatic
flavor effects. This is certainly true for inverted and quasi-degenerate neutrino
mass spectra. Only for normal hierarchy, a reduced washout by an order of magnitude
in the electron flavor can be naturally achieved. This is the reason why we did not
obtain any constraint in that case.

\section{Type III Seesaw}
The Type III seesaw has the same parameters as the Type I, namely a neutrino Yukawa coupling
matrix $\lambda$ and a Majorana mass matrix $M$ which can be taken to be diagonal. The essential
difference is that the seesaw mediators are fermionic $SU(2)$ triplets, with one neutral
and two charged components. Leptogenesis in this model was studied in~\cite{HambyeIII}.
In particular, the washout parameter in Eq.~(\ref{Boltz}) is given by
\begin{equation}
W^{\rm III}_{\alpha}(z)={1\over 4}\sum_i  K^{\rm III}_{i\alpha}
\mathcal{K}_1(z_i) z_i^3,
\end{equation}
where $K^{\rm III}_{i\alpha}\equiv 3\,K_{i\alpha}$ because the three components of the
triplet contribute to the washout.

We now study the constraints on the spectrum of the triplets from washout
arguments. In the case of normal hierarchy and $m_1\ll m_{\rm sol}$,
a pre-existing asymmetry in flavor $e$ will again partially escape the
washout, although not as much as in the Type I case: $\sum_i K^{\rm
III}_{ie}>7$, implying a maximum washout factor of $10^{4}$.
Second, for the same reason as in the Type I case, at least one
of the triplets will decay electrophobically.

For an inverted hierarchy and $m_3\ll m_{\rm sol}$, we have $\sum_i K^{\rm
III}_{i\alpha}\geq 36$. We then obtain the conservative upper bound $M_1<300$~GeV,
substantially more stringent than the corresponding
type I case.

For a quasi-degenerate spectrum, $m_1\simeq m_2\simeq m_3\simeq
0.1$~eV, we obtain
that $\sum_i K^{\rm III}_{i\alpha}\geq 120$. This leads to the upper bound $M_1<170$~GeV.

It is worth pointing out that the LHC has the capability to observe
the triplet fermions up to 1~TeV in five years of operation~\cite{strumia}.
A considerable portion of this allowed range would therefore be in conflict with
high-scale baryogenesis.

\section{Type II Seesaw}\label{sec:typeII}
In the Type II seesaw, neutrino masses are generated by the VEV of the neutral component
of scalar $SU(2)$ triplet $\Delta$. For our purposes, the relevant part of the Lagrangian is
\begin{equation}\label{typeII}
-h_{\alpha \beta} \ell_{L\alpha}^T C {\rm i} \sigma_2 \Delta \ell_{L\beta}- \mu H^T {\rm i} \sigma_2 \Delta^{\dagger}
H +h.c. \, ,
\end{equation}
where $\alpha,\beta=e,\mu,\tau$, and the neutrino mass matrix is given by
\begin{equation}
(m_{\nu})_{\alpha \beta}=h_{\alpha \beta} v_{\Delta}=h_{\alpha \beta}{\mu v^2\over M_{\Delta}^2}.
\end{equation}
It should be noted that only one triplet is able to generate three active neutrino masses, contrary
to the Type I and III cases.

We are interested in the case where $\Delta$ mass in the 100 GeV to TeV range so that it is accessible at LHC~\cite{han}.
We vary the parameter $\mu$ so that the triplet VEV $v_\Delta = \mu v^2/M^2_{\Delta}$ also correspondingly
changes. Note that electroweak $\rho$-parameter constraint implies $v_{\Delta}\leq 1$~GeV. We will always stay far below
this value. Depending on the magnitude of $\mu$, the decay of the $\Delta^{++}$ will either be to a Higgs pair
$\Delta\to H^+H^+$ or to a like-sign dilepton pair $\Delta\to \ell\ell$. Which lepton pair will of course depend on the
neutrino mass hierarchy; specifically, we expect the dominant channels to be $\tau\tau$, $\mu\tau$ and $\mu\mu$
for normal hierarchy and $e\mu$ and $e\tau$ or $ee$ for inverted hierarchy.

Due to the presence of the scalar triplet, which carries a non-zero hypercharge,
the washout is here evaluated solving a set of Boltzmann equations, as presented in~\cite{HambyeII}.
More precisely, coupled equations for the evolution of the asymmetries in the Higgs field, in the scalar triplet,
as well as in each lepton flavor must be solved. Including flavor effects and the most relevant spectator
processes, i.e. the Yukawa interactions~\footnote{We did not include sphaleron effects, which only induce
small corrections.},
and imposing hypercharge
conservation, which allows one to remove one equation, we obtain the following system of equations:
\begin{eqnarray}
{{\rm d} Y_{\ell_{\alpha}}\over {\rm d}z}&=&-2 D_{\Delta} B_{L\alpha}\left(Y_{\Delta}+3z^2 \mathcal{K}_2(z)Y_{\ell_{\alpha}}
\right)\, ,\label{typeIIeq1}\\
{{\rm d} Y_{\Delta}\over {\rm d}z}&=&-2 D_{\Delta}\left[Y_{\Delta} +3z^2\mathcal{K}_2(z)\left(\sum_{\alpha}B_{L\alpha}
Y_{\ell_{\alpha}} \right)\right. \label{typeIIeq2}\\
 && \left.-3 z^2\mathcal{K}_2(z)B_H{2\over 3N_f+5}\left(-2Y_{\Delta}+{8\over 3}\sum_{\alpha} Y_{\ell_{\alpha}}\right)\right]\, ,\nonumber
\end{eqnarray}
where $N_f$ is the number of quark generations, and where we defined
\begin{equation}
D_{\Delta}={\Gamma_{\Delta} \mathcal{K}_1(z)/\mathcal{K}_2(z)\over H z}
\end{equation}
with the total triplet decay rate $\Gamma_{\Delta}={M_{\Delta}\over 16 \pi}\left(\sum_i m_i^2/v_{\Delta}^2
+v_{\Delta}^2M_{\Delta}^2/v^4\right)$. The branching ratios in each flavor $\alpha$ are given by
\begin{equation}
B_{L\alpha}= {\sum_{k} m_k^2 |V_{\alpha k}|^2\over \sum_k m_k^2 +v_{\Delta}^4 M_{\Delta}^2/v^4},
\end{equation}
where $V$ is the PMNS matrix without Majorana phases, and $\sum_{\alpha} B_{L\alpha}\equiv B_L=1-B_H$.

\begin{figure}[t!]
\includegraphics[width=0.3\textwidth,angle=-90]{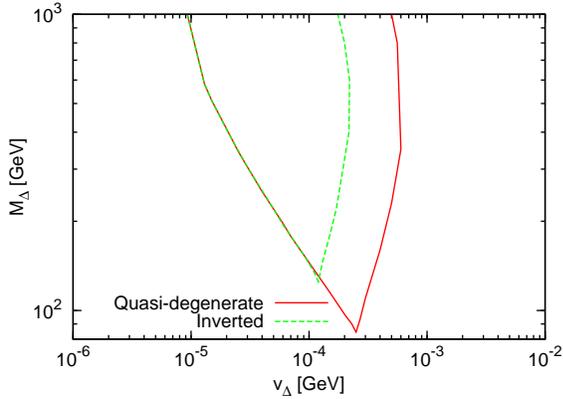}
\caption{Regions in the space $(v_{\Delta},M_{\Delta})$ between the lines
are excluded
because of a washout greater than a factor of $10^{6}$. The cases of inverted
hierarchy with $m_3\ll m_{\rm sol}$ as well as the case
of quasi-degenerate neutrinos ($m_i\simeq 0.1$~eV) are depicted.} \label{fig:2}
\end{figure}
We solve Eqs.~(\ref{typeIIeq1}) and (\ref{typeIIeq2}) with initial asymmetries in the lepton
fields and Higgs fields so as to satisfy hypercharge neutrality\footnote{This is a conservative
choice of initial conditions. An initial asymmetry in the scalar triplet fields would lead to 
more stringent limits.}, and we obtain the region of parameters
$(v_{\Delta},M_{\Delta})$ where the washout is
greater than a factor of $10^{6}$. The result is shown
in Fig.~\ref{fig:2} for the different neutrino mass spectra. It can be seen
that a substantial portion of the parameter space is disfavored by our washout argument in
the case of inverted and quasi-degenerate spectra. The washout for a normal hierarchy is again
suppressed, with no relevant bound.

It is interesting to notice that the system of
equations~(\ref{typeIIeq1}) and (\ref{typeIIeq2}) reduces to one
equation of the form Eq.~(\ref{Boltz}) in the limit $B_L\ll 1$, or
equivalently $v_{\Delta}\gtrsim 10^{-4}$, and when the number of
quark generations is large, $N_f\gg 1$. Indeed, to first order in
small $B_L$, starting with zero initial asymmetry in $\Delta$, from
Eq.~(\ref{typeIIeq2}) it follows that $Y_{\Delta}$ remains
negligible, and therefore it can be neglected in
Eq.~(\ref{typeIIeq1}). In that case, we obtain that the washout
function in Eq.~(\ref{Boltz}) is given by
\begin{equation}\label{washoutIIzero}
W^{\rm II}_{\alpha}(z)= 3{v^2\sum_{k}m_k^2 |V_{\alpha k}|^2 \over
m_{\star}v_{\Delta}^2 M_{\Delta}} \mathcal{K}_1(z) z^3 \, .
\end{equation}
If the branching ratios in each lepton flavor $B_{L\alpha}$ are
similar, as it is the case for a quasi-degenerate spectrum, it is
possible to go to second order in small $B_L$ and still find a
simple equation of the form Eq.~(\ref{Boltz}). From
Eq.~(\ref{typeIIeq2}) it follows that $Y_{\Delta}$ will remain small
if $Y_{\Delta}=-3z^2\mathcal{K}_2(z)\sum_{\alpha}B_{L\alpha}
Y_{\ell_{\alpha}}$. Strictly speaking this is only consistent if
$B_L\ll 1$. Replacing in Eq.~(\ref{typeIIeq1}), we obtain that the
washout function in Eq.~(\ref{Boltz}) is given by
\begin{equation}\label{washoutII}
W^{\rm II}_{\alpha}(z)= 3{M_{\Delta}v_{\Delta}^2v^2\sum_{k}m_k^2
|V_{\alpha k}|^2 \over m_{\star}(v^4\sum_k m_k^2+v_{\Delta}^4
M_{\Delta}^2)} \mathcal{K}_1(z) z^3 \, .
\end{equation}
The approximations~(\ref{washoutIIzero}) and (\ref{washoutII}) were
derived in the limit of a large number of quark generations, $N_f\gg
1$. For $N_f=3$, we find that using Eq.~(\ref{washoutIIzero}) or
(\ref{washoutII}) to compute the exclusion region shown in
Fig.~\ref{fig:2} would imply a change of less than 10\% in the
region $v_{\Delta}\gtrsim 10^{-4}$. Moreover, even though the
approximation~(\ref{washoutII}) is not strictly speaking valid for
$B_L\sim 1$, or equivalently $v_{\Delta}\lesssim 10^{-4}$, it is
also relatively good in that range; the bounds presented in
Fig.~\ref{fig:2} would become less stringent by no more than 30\%.
Note that the Yukawa interactions which redistribute the Higgs
asymmetry among all SM particles, and which are at the origin of the
factor ${2\over 3N_f+5}$ in Eq.~(\ref{typeIIeq2}), are absolutely
essential to get the correct result. Using strictly the equations
proposed in~\cite{HambyeII} adding only flavor effects, and
therefore neglecting these Yukawa interactions, leads to an
overestimation of the washout by up to 12 orders of magnitude! The
bounds presented in Fig.~\ref{fig:2} would then become more
stringent in the region $v_{\Delta}\lesssim 10^{-4}$ by up to a
factor of 2.

\section{Conclusion}

We studied in detail the consequences of weak-scale seesaw mechanisms on any
pre-existing baryon asymmetry. If it is not to be efficiently washed out, we found
correlations between the pattern of neutrino masses and the spectrum of new particles
at the weak scale. For type I seesaw models with a TeV scale $Z'$ coupled to SM
neutrinos, we found that for a normal neutrino mass hierarchy, at least one of the RH neutrinos must be
`electrophobic', decaying with a strong preference into final states with muons and tauons
rather than electrons. For
inverted or quasi-degenerate mass patterns, on the other hand, we found upper bounds on the mass of at least one
RH neutrino. In particular, for an inverted mass hierarchy, this bound is 1 TeV, while the corresponding upper
limit in the quasi-degenerate case is 300 GeV. Similar results hold in type III seesaw models, albeit with somewhat more
stringent bounds. For the Type II seesaw case with a weak scale $SU(2)$ triplet Higgs, we again found that an interesting
range of Higgs triplet masses is disallowed by these considerations.

\begin{acknowledgments}
The works of ZC and RNM are supported by the NSF under grants PHY-0801323 and PHY-0652363 respectively.
\end{acknowledgments}

\end{document}